\documentstyle[aps,prb,multicol,graphicx]{revtex}
\input{epsf.sty}
\begin{document}
%\bibliographystyle{klunum}
%\tableofcontents
%\begin{article}
\def\alt{\lsim}
\def\agt{\gsim}
\title{Quantum interferometry with electrons:\\
outstanding challenges}
\author{Yuval Gefen}
\address{Department of Condensed Matter Physics, The Weizmann
Institute of Science, Rehovot 76100, Israel\\ }
% PLEASE NOTE: NO ABSTRACT IS REQUIRED
%KEYWORDS: Please include 15-20 your KEYWORDS after % sign.
%They will be indexed on compiling the entire volume
%\end{opening}
%\begin{document}
\maketitle
\tableofcontents
\section{Introduction and Perspectives}
Much of the effort in studying  mesoscopic systems was directed
towards the analysis of thermodynamic or transport properties {\it
per se}. However, measurements (and, subsequently, the analysis)
of such quantities can teach us  a lot about the  {\it  phase} of
the wavefunctions involved. Evidently, one needs to explicitly
incorporate quantum interference into such measurements  to allow
for the  analysis of the quantum phase. Earlier interferometry
experiments  in mesoscopic conductors focused on various aspects
of  Aharonov-Bohm (AB)  \cite{Aharonov} oscillations vis-a-vis
transport \cite{Sharvin,Webb1,Webb2,Webb3} or thermodynamics
(persistent currents and orbital magnetism
\cite{Levi-Bouchiat,Webb,Mailly,Mohanty}), initially interpreting
the data within the framework of single electron physics. This
indeed led to an impressive number of novel and, at times,
unexpected effects. Over the past few years it has become clear,
though, that the physically motivated, yet naive, picture  of
independent electrons does not suffice to account for the
important details that  have emerged in the course of the
experimental work. Moreover, further theoretical analysis
suggested that the presence of  electron-electron interaction may
indeed give rise to novel important physics.

\begin{figure}
\epsfxsize = 0.8 \textwidth
\centerline{\epsfbox{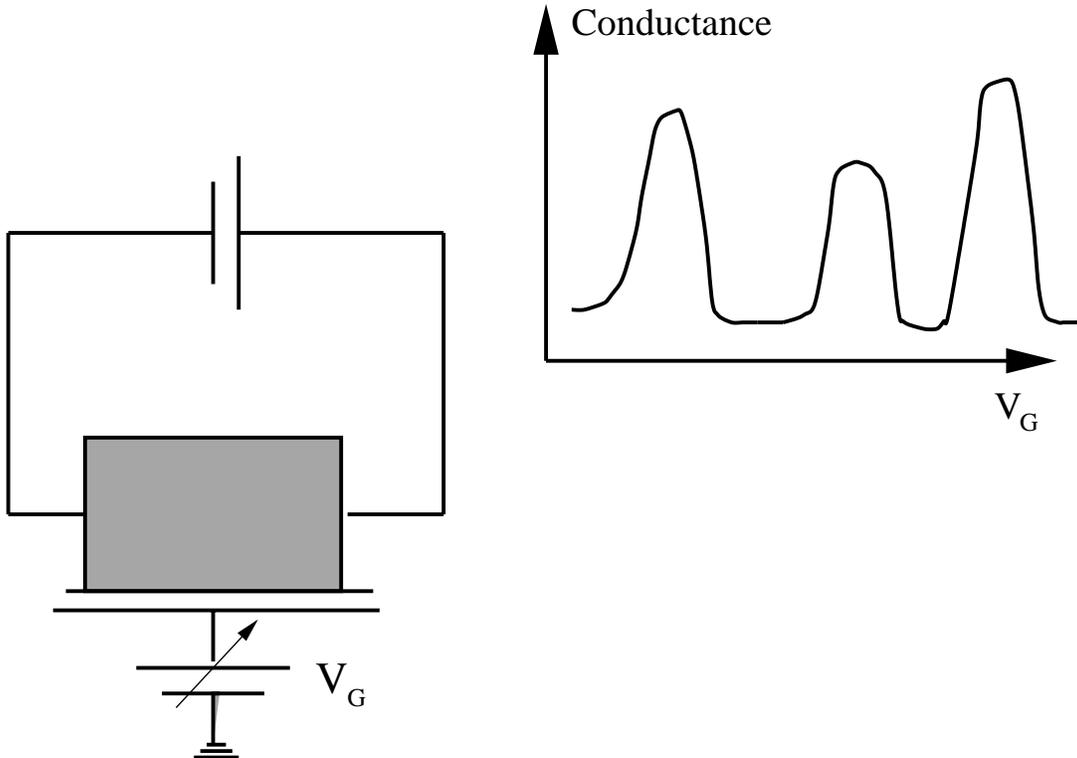}}
\caption{A schematic equivalent circuit of a quantum dot. Inset: The linear
conductance through the dot  as function of the gate voltage $V_G$ (schematic).}
\label{fig1}
\end{figure}

Thus  the new generation of interference experiments
\cite{Yacoby,Schuster,Ji1,Ji2,Kouwenhoven} focused on setups where
the role of  e-e  interaction is emphasized and can be controlled.
(Here we leave  aside the very interesting issue of thermodynamics
of mesoscopic systems).  The obvious choice is to incorporate
quantum dots (QDs) \cite{Kouwenhoven97,Alhassid,Aleiner_B_G} in
the interferometers. For QDs which are ``closed'' (i.e.
unconnected to external leads) the electron-electron interaction
may be modeled by a $0$-mode capacitive energy  term, that is  an
interaction term which does not have any spatial dependence.
Higher, space-dependent, interaction modes are smaller in powers
of the inverse diemsnionless conductance of the uncoupled dot
\cite{KG_unpublished,Blanter,AGKL,Blanter_Mirlin}, $1/g$.
Fig.~\ref{fig1}  depicts  schematically the conductance through a
QD as the applied gate voltage, $V_G$,  is varied. Coulomb peaks
appear at values of $V_G$ for which the energy of the entire
system  (i.~e., the QD  and the reservoirs) is ( nearly)
insensitive to the removal/addition of an electron from the leads.
Hereafter we denote the width of the Coulomb peaks by $\tilde
\Gamma$. Such a quantum dot can now be  incorporated into an AB
interferometer. The two obvious parameters to vary are the
enclosed AB flux, $\Phi$, and the gate voltage on the QD. The
former is parametrized  by $\varphi \equiv 2\pi \Phi/\Phi_0$,
where $\Phi_0 \equiv e/hc$ is the flux quantum. Evidently,  one
may study the dependence on other important parameters, such as
the temperature $k_B T$, the dimensionless conductance of the
uncoupled dot $g$, the coupling strength of the QD to the leads
$\Gamma$ etc.

%%%%%%%%%%%%%%%%%%%%%%%%%%%%%%%%%%%%%%%%%%%
%%%%%% figure(fig1.eps) goes here %%%%%%
%%%%%%%%%%%%%%%%%%%%%%%%%%%%%%%%%%%%%%%%%%%%
%%%%%%%%%%%%%%%%%%%%%%%%%%%%%%%%%%%%%%%%%%%%

The quantum dot interferometry experiments yielded a rich set of
results. Most of them came initially as a surpirse, but further
analysis  provided satisfactory  explanations to the observed
effects  within single particle theories or otherwise  the
orthodox theory of Coulomb blockade. Certain  effects, though,
turned out to be more puzzling, initiating an extensive
theoretical effort which is yet to prove fruitful.  Some of the
puzzling results even went unnoticed. I  shall not attempt  to
present here a comprehensive review of the physics of coherence
and interferometry on mesoscopic scales, nor shall I review all
the relevant ( and important) literature. Instead, I provide an
updated compendium of the experimental results, the dilemmas they
present, and some theoretical perspectives concerning  the
attempts to address this physics. In fact, the present  overview
can be regarded as  a shopping list of the present challenges in
this field.  Due to the limited scope of this presentation I will
not address the  very low temperature  limit, where {\it Kondo
physics} comes to play. Interesting effects are expected in that
context, cf. e.g. Refs.~
\cite{Ji1,Ji2,Oreg,Chiappe,Hofstetter,Affleck}. I also note that
the magnetic fields discussed here are sufficiently weak to ignore
spin-related Zeeman splitting. Likewise, the systems considered
are always small (or narrow) enough for the quantum Hall effect
not to show up.

The outline of this paper is the following. In the next section I
review some basic facts concerning AB interferometry in electronic
 systems, including a reference to symmetries at equilibrium (Onsager) and away
 from equilibrium, a brief discussion of partial coherence, and a simple
 demonstration of the breakdown of the Landauer formula in the presence of interaction.
In Section \ref{Phase_locking} I discuss the  phenomenon of phase
locking (and a scenario for its breakdown). Section
\ref{Transmission_phase} addresses the issue of the transmission
phase correlations. In Section \ref{Asymmetry} we discuss the
asymmetric features of the coherent AB amplitude. Finally, in
Section \ref{Width} I comment on the dilemmas involving the width
of the Coulomb peaks as well as that of the phase lapses.

This short overview is largely based on past and present
collaborations. Parts of this manuscript consist of results
obtained recently in collaboration with D,~E.  Feldman, J.
K\"onig, Y. Oreg and  A. Silva.

\section{Aharonov-Bohm interferometry in electronic systems: some basics}
\label{AB_general} Earlier on in the development of the field of
{\it Mesoscopics} it became clear that AB interferometry is a most
useful tool. This, of course, has to do with the fact that
electrons are charged particles, and therefore the electric
current is (minimally) coupled to the vector potential, {\bf A},
generated by a Aharonov-Bohm flux. The vector field {\bf A}
influences  the phase of the electron and thus affects the outcome
of interference experiments. The first intriguing effect that has
been noticed in that context was  that while the periodicity (say,
of the conductance) under the AB flux should, in principle, be
$\delta \varphi =1$  \cite{GIA1,GIA2},  the AB periodicity of a
flux threading a (dirty) conducting cylinder was found to be
$\delta \varphi ={1 \over  2}$ \cite{AAS}, a result which was
confirmed in experiment \cite{Sharvin}. While the latter
periodicity was certainly in line with the Byers-Yang  theorem
\cite{BY}, stating that equilibrium (hence linear-response)
observables should be periodic under $\varphi \rightarrow
\varphi+1$, the effect of period halving came as a surprise. It
was later understood \cite{Gefen84} that period halving was the
result of ensemble averaging \cite{Murat,Stone_Imry}, and that in
the absence of such averaging the fundamental  period should be
$1$ rather than ${ 1\over 2}$. This indeed was found to be in line
with experimental observations \cite{Webb1,Webb2}. Other exciting,
interferometry-related, effects followed (such as persistent
currents, orbital magnetism,
flux-dependent ``universal'' conductance fluctuations).\\

{\bf Phase locking}\\

\begin{figure}
\epsfxsize = 0.5 \textwidth
\centerline{\epsfbox{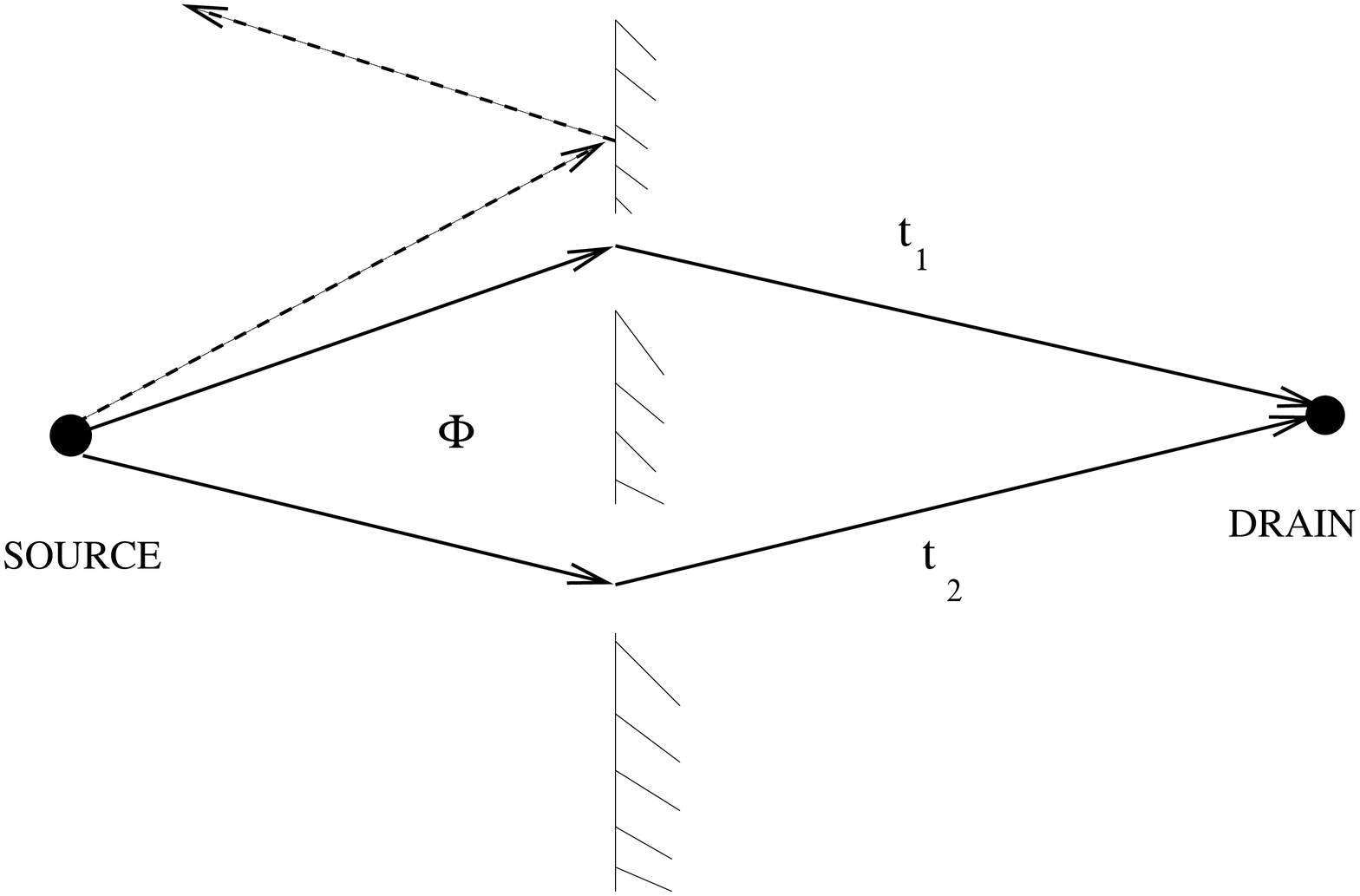}}
\caption{Setup for a two-slit interferometry.The dashed line
represents a trajectory of an electron absorbed by a ``far-gate''.}
\label{fig2}
\end{figure}

One concept that  emerged from those early studies was the
phenomenon of {\it phase locking}
\cite{Buettiker86,refGIA1,ImryBook}. To understand this effect let
us first consider  a two-slit experiment,  depicted in
Fig.~\ref{fig2}, for which phase locking is not satisfied.

%%%%%%%%%%%%%%%%%%%%%%%%%%%%%%%%%%%%%%%%%%%
%%%%%% figure(fig2.eps) goes here %%%%%%
%%%%%%%%%%%%%%%%%%%%%%%%%%%%%%%%%%%%%%%%%%%%
%%%%%%%%%%%%%%%%%%%%%%%%%%%%%%%%%%%%%%%%%%%%

 A single electron physics is assumed. The electron
is emitted from the source and may be absorbed by either the drain, the source,
or be ``lost''  (i.e. absorbed by any of the other ``far gates''). This motivates
later reference to this setup as {\it open geometry}. The
partial amplitudes for the electron to be transmitted through slits 1 or 2
( and eventually be absorbed by the drain) are $t_1$ and $t_2$ respectively, with

\begin{eqnarray}
\label{partial_transmission_amplitude}
t_m=|t_m| \exp(i \alpha_m)\,\, ,
\end{eqnarray}
with $m=1,2$. In the presence of an AB flux  the  partial transmission amplitudes
assume  additional flux induced, gauge dependent phases,

\begin{eqnarray}\label{partial_AB_phases}
\alpha_m \rightarrow \alpha_m + 2 \pi \varphi_m\,\, .
\end{eqnarray}

The relative phase of the two trajectories described by these two partial
amplitudes is

\begin{eqnarray}
\label{relative_AB_phase}
\alpha + 2\pi \varphi \equiv
\alpha_1 - \alpha_2 +2\pi \varphi_1 -  2\pi \varphi_2 \,\, ,
\end{eqnarray}
where $\varphi \equiv \varphi_1-\varphi_2$ is the gauge invariant AB phase.
The total transmission amplitude, $t_{total}=t_1+t_2$.
Employing the Landauer formula
\cite{Landauer,Buettiker86}, the total transmission {\it probability} is
given by
\begin{eqnarray}
\label{Landauer}
T_{tr}=|t_{total}|^2\,,
\end{eqnarray}
where $t_{total}$ is the total transmission amplitude. For our geometry
it follows that
\begin{eqnarray}
\label{ltp}
T_{tr}=|t_1|^2+|t_2|^2+2|t_1 t_2| \cos(\alpha +2\pi
\varphi) \,\, .
\end{eqnarray}
The flux sensitive interference term  $2|t_1 t_2| \cos(\alpha
+2\pi\varphi)$ is evidently periodic  in $\varphi$ (flux) with a
period $1$. The asymmetry of  the AB signal  with respect  to
$\varphi=0$ is due to the appearance of the orbital phase
$\alpha$. It is quite suggestive to refer to a situation where
this phase shift disappears (e.g. due to some underlying symmetry,
see below)
as {\it phase locking}.\\

{\bf Partial coherence and visibility}\\

The  above expression for the total transmission probability
through the double-slit configuration  was derived under the
conditions of full coherence. The conductance is related to the
transmission probability through

\begin{eqnarray}
\label{conductance_transmission}
{\cal G}={e^2 \over h}T_{tr}\,\,.
\end{eqnarray}

This relation holds for interacting systems as well. To check
whether there is some degree of cohernece in the system is a
relatively easy task. We only need to note that there are certain
values of the flux for which the total transmission is smaller
than the sum of the individual transmissions through each channel,
i.e., $T_{tr} < |t_1|^2 + |t_2|^2$. To assert that there is {\it
full coherence} in the system is  a more demanding task. Consider
the expression for the total transmission, Eq.\ref{ltp}. For
$|t_1| \neq |t_2|$ the AB amplitude is smaller than the
flux-averaged signal. Referring to the conductance we can write
the above inequality  as
\begin{eqnarray}\label{partial_coherence}
\left\langle {\cal G} \right\rangle_{\varphi}={e^2 \over
h}(|t_1|^2+|t_2|^2) >{\cal G}^{AB}=2 {e^2 \over h} |t_1 t_2|> 0
\,\, ,
\end{eqnarray}
with $\left\langle {\cal G} \right\rangle_{\varphi}$ being the
flux-averaged conductance, ${\cal G}^{AB}$ is the amplitude of the
(periodic) flux dependent term. It is therefore convenient to
define the {\it visibility}, ${\cal V}$, of such an interferometer
\begin{eqnarray}\label{visibility}
{\cal V} \equiv
{\cal G}^{AB} /  \left\langle {\cal G} \right\rangle_{\varphi}  \,\, .
\end{eqnarray}
There could be 3 different reasons for the visibility to be smaller than one:
\\ (i) The transmission  is fully coherent, yet the  transmissions  through the
two interferometer arms are asymmetric--one  arm  transmits better
than the other. This is the scenario outlined above. Evidently
when the two interferometer's arms are symmetric,$|t_1| = |t_2|$,
${\cal V}=1$. \\(ii) There are several transmission channels
through each arm, each carrying its own orbital (and possibly AB)
phase. It follows that the conditions for destructive interference
are different among the different channels, and may not be
satisfied simultaneously. \\(iii) The transmitted electrons are
coupled to other degrees of freedom \cite{Koenig01,Mello} which
give rise to dephasing, setting the stage for {\it partial
coherence}.

The first two scenarios are fundamentally different from the third one, as they
correspond to full coherence (although the observed visibility may be
smaller than unity). Indeed,  full coherence implies that {\it in principle}
it is possible to tune or modify the parameters of   one of the  arms
( ``the reference  arm''), and vary the flux such that  full destructive
interference (${\cal V}=1$) is obtained.

From the theoretical point of view there are two main approaches
for  incorporating dephasing processes in  interferometry devices.
The first one \cite{Buettiker85,Riedel} is phenomenological. One
starts with writing down a scattering theory formalism for the
problem at hand ({\it a-la} Landauer),  and then adding
``dephasing reservoirs'' which  absorb and emit electrons, usually
without modifying the current. What a dephasing reservoir does is
to erase any phase memory of the electrons that go through.
Operationally one describes the scattering  into/out of the
dephasing reservoir by assigning some complex amplitude to such a
process, with a phase $\theta$. Once an observable (e.g. the
transmission probability through the device) is calculated, an
average over $\theta$ is taken. Various generalizations of this
approach are possible, e.g. the introduction of numerous, weakly
coupled reservoirs along the transmission line which mimicks
gradual, continuous dephasing processes \cite{Buettiker85,Riedel}.
There are certain caveats with this procedure: The dephasing agent
here is completely classical. This means that in the process of
dephasing energy may be pumped into to electronic system
\cite{Gefen_Schoen}. Also, subtle quantum effects and correlations
are ignored in this approach. Finally, the averaging over the
phase  $\theta$ does not commute with the time reversal operation;
performing this averaging and then time reversing the problem may
present problems with  unitarity \cite{Riedel}. The second
approach to the introduction of dephasing is microscopic
 One can classify the dephasing agents according to
whether the coupling term in the Hamiltonian does or does not
commute with the Hamiltonian of the uncoupled reservoir
 \cite{Shimshoni_Gefen}. The physics that emerges is elaborate and
not yet fully exhausted.\\

{\bf Two-terminal vs. multi-terminal geometries}\\

\begin{figure}
\epsfxsize = 0.5 \textwidth
\centerline{\epsfbox{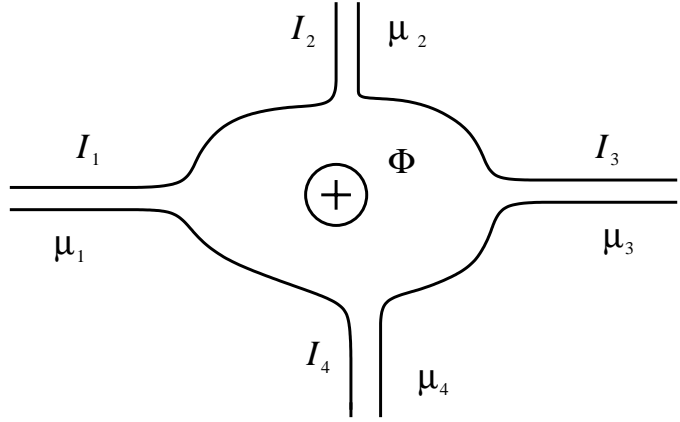}}
\caption{A 4-terminal AB interferometer.}
\label{fig3}
\end{figure}

As we have seen above, owing to the different orbital phases of
the two interferometer's arms,  the  AB signal is, in general,
asymmetric with respect to $\varphi=0$, i.e., no phase locking
takes place. This is the case for the open geometry depicted in
Fig.~\ref{fig2}, and similarly  for a multiple-terminal setup,
Fig.~\ref{fig3}.

%%%%%%%%%%%%%%%%%%%%%%%%%%%%%%%%%%%%%%%%%%%
%%%%%% figure(fig3.eps) goes here %%%%%%
%%%%%%%%%%%%%%%%%%%%%%%%%%%%%%%%%%%%%%%%%%%%
%%%%%%%%%%%%%%%%%%%%%%%%%%%%%%%%%%%%%%%%%%%%

In contradistinction, for  a two-terminal
geometry  one expects  phase locking to take place. We note
that in such a geometry an electron leaving the source may, eventually, be either
reflected back to the source or transmitted to the drain ( unlike in a
multi-terminal geometry or in an open geometry where the electron may end up
in one of the other ``gate''  terminals). The total transmission and the reflection
probabilities satisfies then

\begin{eqnarray}
\label{unitarity}
 T_{tr}+R_{ref}=1  \,\, .
\end{eqnarray}
Employing Eqs. \ref{conductance_transmission} and \ref{unitarity}
the flux dependence of ${\cal G}$ can be fully deduced from
$R_{ref}$, the latter obtained by taking the square modulus of the
sum of the self-returning amplitudes. To accomplish this task we
pretend that those amplitudes can be evaluated through an infinite
set of semiclassical trajectories  which we denote by
$\{A_j^{(n)}\}$, \cite {footnote_A}.  This notation is rather
symbolic, pretending that there is a countable set of relevant
partial amplitudes. We classify the various partial amplitudes
according to their winding number, $n$, and a running index $j$
within each class of $n$. Each of these partial amplitudes
includes an orbital phase which we denote by $\alpha_j^{(n)}$.
Furthermore, we need to assign an AB phase (which  depends on the
winding number only), $\varphi^{(n)} \equiv 2 \pi n \varphi$. Thus
\begin{eqnarray}
\label{partial_amplitude} A_j^{(n)}=|A_j^{(n)}|\exp(i
\alpha_j^{(n)})\exp(i\varphi^{(n)})\,\,.
\end{eqnarray}
Let   $A_j^{(n)}$ and  $A_j^{(-n)}$ represent two trajectories which  are
mutually time reversed. It follows that
\begin{eqnarray}&& \label{time_reversed_phases} A_j^{(n)}=A_j^{(-n)} \,\, ,
\nonumber
\\&& \varphi^{(n)}=-\varphi^{(-n)} \,\,.
\end{eqnarray}
The reflection probability is then given by
\begin{eqnarray}
\label{total_reflection}
R_{ref}=r \times r^* = \left[ \sum_{j,n}
A_j^{(n)} \right] \times \left[ \sum_{j,n} A_j^{(n)} \right]^*
\,\,.
\end{eqnarray}
Performing this multiplication we obtain several distinct types of terms:

(i)\,  {\bf ''diagonal terms''}, arising from the multiplication
of $A_j^{(n)}$ by $\left[ A_k^{(n)} \right]^*$ (same winding
number, $j,k$ are general). These sample specific terms  amount to
a flux independent  contribution. In particular, the product of
the partial amplitude  $A_j^{(n)}$ with its complex conjugate rids
of the (sample specific) orbital phase. Adding together the
diagonal contributions of $A_j^{(n)}$ and $A_j^{(-n)}$ one obtains
\begin{eqnarray}
\label{diagonal_contributions} |A_j^{(n)}|^2+ |A_j^{(-n)*}|^2=
2|A_j^{(n)}|^2  \,\,.
\end{eqnarray}

(ii)\,  {\bf ''time reversed'' terms},  arising  from the product
of $A_j^{(n)}$ and $\left[ A_k^{(-n)} \right]^*$ with $j=k$ (we
will be concerned with time-reversed pairs; $j \neq k $ pairs
possess sample specific   orbital phases; such terms contribute to
the statistical fluctuations, in much the same way as the
cross-terms of type (iii)). By Eq.\ref{time_reversed_phases} the
orbital phase of a time-reversed pair cancels out, and one is left
with a flux-dependent phase only. Adding  up two related reversed
pairs one obtains
\begin{eqnarray}\label{time_reversed_pairs}
A_j^{(n)} \times A_j^{(-n)*}+A_j^{(-n)} \times A_j^{(n)*}=
2|A_j^{(n)}|^2 \cos (4 \pi n  \varphi)  \,\,.
\end{eqnarray}

(iii)\,  {\bf ''cross terms''}. These are all the rest. Their
magnitude is sample specific, their  (orbital)  phase is sample
specific (hence strongly fluctuating), and in general they are
flux dependent.

Detailed analysis of these terms may yield a wide spectrum of
effects, basically the entire single-electron mesoscopics in a
nutshell, including the effects of period-halving, negative
magneto-resistance at weak magnetic fields, conductance
fluctuations and more. This is beyond the scope of the present
analysis. What we would like to stress here is the emergence of
phase locking. This can be easily seen for the ensemble averaged
problem. The terms which survive ensemble averaging  are given by
Eqs.(\ref{diagonal_contributions}, \ref{time_reversed_pairs})
(summation over $j$  is implied); all the other, strongly
fluctuating, terms average to zero. Symmetry with respect to
$\varphi=0$ is evident. Phase locking, though, is a more robust
phenomenon, valid on  the level of sample-specific observables. We
only need to note that when performing the multiplication implied
by Eq.\ref{total_reflection} we add up the following terms
together: $A_k^{(n)} \times  A_j^{(m)*}+A_k^{(n)*} \times
A_j^{(m)}+
A_k^{(-n)} \times  A_j^{(-m)*}+ A_k^{(-n)*} \times  A_j^{(-m)*}=\\
4|A_j^{(m)}A_k^{(n)}|\cos(\alpha_j^{(m)}-\alpha_k^{(n)})\cos((n-m)\varphi)$.
Phase locking is then manifest.

Dephasing or inelastic relaxation processes may suppress the
coherent transmission and  reflection amplitudes (giving rise to
an incoherent background)-- more generally they will suppress the
related single-particle Green function. It is important to notice,
though, that such processes do not destroy the phase locking
symmetry. A mechanism for breaking down this symmetry is discussed
in Section  \ref{Phase_locking}.

Careful examination of our analysis of the noninteracting system
reveals that phase locking is satisfied within any infinitesimal
energy window (we have been dealing with partial amplitudes each
possessing a certain energy). In that case the validity of phase
locking ( for the two-terminal geometry) is independent of the
electron energy distribution function, and it extends beyond
linear response. This is not the case in general, where
electron-electron interactions render the notion of a ``sharply
defined single electron energy'' inadequate. One may still prove
the validity of phase locking assuming that the reservoirs at the
boundaries of the system (e.g. the source and the drain) are
described by a gas of non-interacting electrons . This proof,
originally proposed by B\"uttiker \cite{Buettiker86}, relies on
the general Onsager relations. The fact that a finite external
voltage bias can destroy phase locking has been nicely
demonstrated by Bruder, Fazio and Schoeller \cite {Bruder}.

\begin{figure}
\epsfxsize = 0.5
\textwidth\centerline{\epsfbox{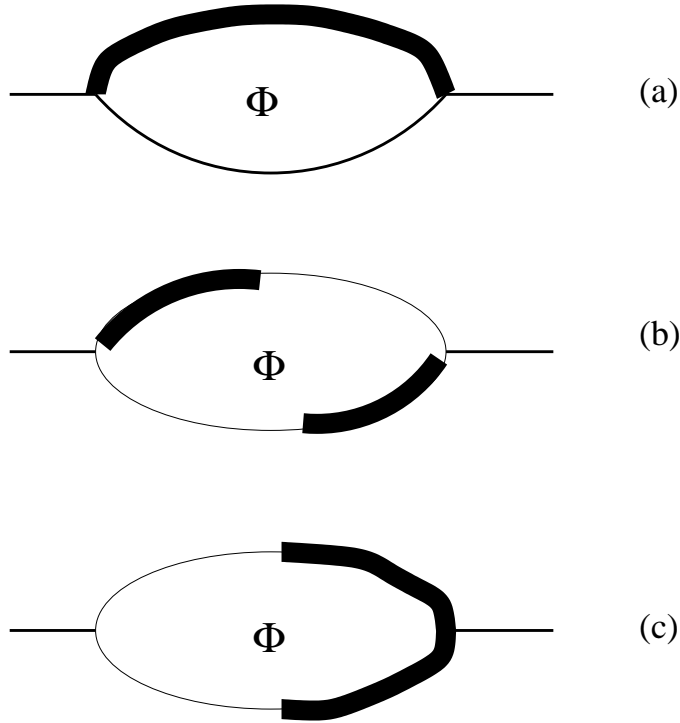}} \caption{Two
terminal AB interferometers with different spatial symmetry.
  Different lines represent different lead geometries/impurity configurations.
  (a) possesses mirror symmetry with respect to a vertical axis, (b) possesses
  a point symmetry:  invariance
  under rotation at angle $\pi$, and (c) has a mirror symmetry with respect to a
  horizontal axis. The chemical potentials of the left/right
  reservoirs are indicated.
  In cases  (b) and (c) phase locking is preserved even at finite bias.}
\label{figsymmetries}
\end{figure}

It should be noted, though, that even away from equilibrium, when
linear response does not apply, there may still be certain
symmetries governing the behavior of the conductance. K\"onig and
Gefen \cite{Koenig01} noted the connection between the spatial
symmetries of the underlying setup and the symmetries of the
transport coefficients. In   Fig.~\ref{figsymmetries} we depict
three different cases in which the system has a distinct spatial
symmetries \cite{Koenig01}.
\newpage
%%%%%%%%%%%%%%%%%%%%%%%%%%%%%%%%%%%%%%%%%%%%%%%
%%%%%%%%%%%%%%%%%%%%%%%%%%%%%%%%%%%%%%%%%%%%%%%
%%%%%%%%%%% figsymmetries %%%%%%%%%%%%%%%%%%%%%
%%%%%%%%%%%%%%%%%%%%%%%%%%%%%%%%%%%%%%%%%%%%%%%%
%%%%%%%%%%%%%%%%%%%%%%%%%%%%%%%%%%%%%%%%%%%%%%%%
The general relation for all two-terminal setups
\begin{eqnarray}
   {\partial I (V,\varphi) \over \partial V} =
        {\partial I (-V,-\varphi) \over \partial V} \, ,
\label{general_symmetry}
\end{eqnarray}
where $V$ is the  applied bias, yields as a direct consequence the
Onsager relation
\begin{eqnarray}
   {\partial I (\varphi) \over \partial V} \bigg|_{V=0} =
        {\partial I (-\varphi) \over \partial V} \bigg|_{V=0}
\label{onsager}
\end{eqnarray}
which leads to  phase locking in linear response.
Fig.~\ref{figsymmetries}a represents a system with  a mirror
symmetry with respect to its vertical axis. One clearly can
reverse the direction of the bias {\it and} the sign of the
AB-flux leaving the magnitude of the current unchanged. The
resulting equation is
\begin{eqnarray}
   {\partial I (V,\varphi) \over \partial V} =
        {\partial I (-V,-\varphi) \over \partial V}
\label{symmetry 1}
\end{eqnarray}
which coincides with Eq~(\ref{general_symmetry}).
Fig.~\ref{figsymmetries}b represents a point symmetry: rotation at
angle $\pi$ with respect to the center. The resulting invariance
is expressed through
\begin{eqnarray}
   {\partial I (V,\varphi) \over \partial V} =
        {\partial I (-V,\varphi) \over \partial V} \, .
\label{symmetry 2}
\end{eqnarray}
Finally, Fig.~\ref{figsymmetries}c depicts a mirror symmetry with
respect to a horizontal axis, leading to the equation
\begin{eqnarray}
   {\partial I (V,\varphi) \over \partial V} =
        {\partial I (V,-\varphi) \over \partial V} \, .
\label{symmetry 3}
\end{eqnarray}
In the two latter cases phase locking symmetry is satisfied
\cite{Koenig01}. It either follows directly or after making use of
Eq.~(\ref{general_symmetry}). Here phase locking  is a consequence
of spatial symmetry. In the first case
(Fig.~\ref{figsymmetries}a), or in the absence of any particular
spatial symmetry, breaking of phase locking occurs at finite bias
voltages.\\

{\bf Watch the Landauer formula for interacting systems}\\

The Landauer formula provides a convenient framework to study the
conductance of specific setups,  relating the transmission
amplitude through the system to the total transmission
probability, cf.
Eqs.~(\ref{Landauer},\ref{conductance_transmission}). Things are
not as simple when it comes to an interacting system. In that case
any given electron interacts with other degrees of freedom ( e.g.
other electrons), and its energy  is not conserved-- one needs to
resort to a  many-particle, continuous energy-channel scheme (cf.
\cite{Mello,ImryBook}.  The Landauer formula has indeed been
generalized  employing  Green's function techniques for
interacting systems \cite{Meir92,Koenig96,Koenig99}. To
demonstrate the failure of the naive Landauer formula, and to
relate it to the concepts of partial coherence and decoherence, we
present here a toy model which captures these themes
\cite{Koenig01}. Let us consider a single-level QD with level
energy $\epsilon$, measured from the Fermi energy of the leads.
The Hamiltonian
\begin{eqnarray}
\label{QD_Hamiltonian}
H=H_L+H_R+H_D+H_T
\end{eqnarray}
consists of
$H_r = \sum_{k \sigma} \epsilon_{kr} a^\dagger_{k\sigma r}a_{k\sigma r}$
for the left and right lead, $r=L/R$.
The isolated dot is described by
$H_D = \epsilon \sum_\sigma n_\sigma + U n_\uparrow n_\downarrow$, where
$n_\sigma = c^\dagger_\sigma c_\sigma$, and $H_T = \sum_{k \sigma r} (t_r
a^\dagger_{k\sigma r} c_\sigma + {\rm H.c.})$ models tunneling between dot
and leads (we neglect the energy dependence of the tunnel matrix elements
$t_{L/R}$).
Due to tunneling the dot level acquires a finite line width
$\Gamma=\Gamma_L+\Gamma_R$ with $\Gamma_{L/R}=2\pi|t_{L/R}|^2N_{L/R}$ where
$N_{L/R}$ is the density of states in the leads.
The electron-electron interaction is accounted for by the charging energy
$U=2E_C$ for double occupancy.
To keep the discussion simple we choose
$U=\infty$ for the  QD.

As was  discussed above, a contribution to the transport through a
QD is identified as fully coherent if by adding  a reference
trajectory fully destructive interference can be achieved.
Interaction of the dot electrons with an external bath (e.g.
phonons) destroys coherence since interference with a reference
trajectory is no longer possible: the transmitted electron has
changed its state or, equivalently \cite{Stern90}, a trace in the
environment is left. One possible mechanism for suppressing
coherence in {\it interacting} QDs is   flipping the spin  of both
 the transmitted electron and the QD  \cite{Footnote2}.

\begin{figure}
\epsfxsize = 0.5
\textwidth\centerline{\epsfbox{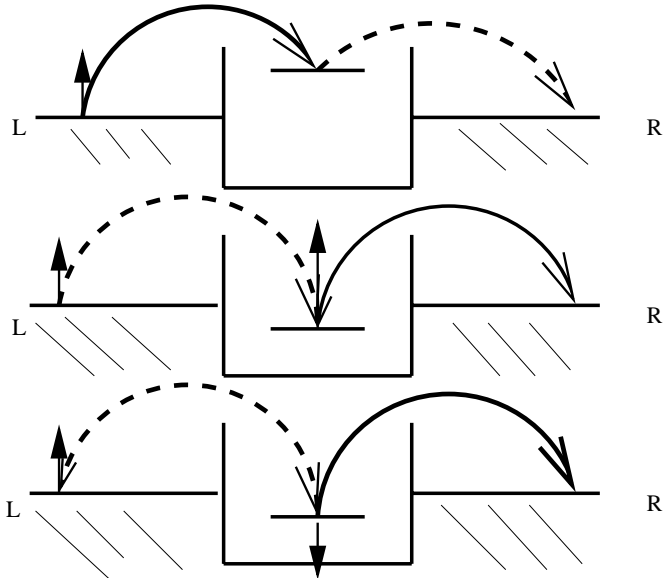}}
\caption{Cotunneling processes for $U=\infty$.
        The solid line indicates the process that happens first, the dashed
        line -- the process that occurs afterwards.
        Double occupancy of the QD in the initial, intermediate, or final
        state is prohibited due to the infinite charging energy.}
\label{figcotunneling}
\end{figure}

Away from resonance, $|\epsilon| \gg k_BT, \Gamma$, transport is dominated by
cotunneling.
There are three different types of cotunneling processes (for $U=\infty$):\\
(i) an electron enters the QD, leading to a virtual occupancy, and then
leaves it to the other side.\\
(ii) an electron leaves the QD, and an electron with the same spin enters.\\
(iii) an electron leaves the QD, and an electron with opposite spin enters.
\\
These three processes are shown schematically in
Fig.~\ref{figcotunneling}.

%%%%%%%%%%%%%%%%%%%%%%%%%%%%%%%%%%%%%%%%%%%%%%%%
%%%%%%%%%%%%%%%%%%%%%%%%%%%%%%%%%%%%%%%%%%%%%%%%
%%%%%%%%%%% figcotunneling %%%%%%%%%%%%%%%%%%%%%
%%%%%%%%%%%%%%%%%%%%%%%%%%%%%%%%%%%%%%%%%%%%%%%%%%%%%%
%%%%%%%%%%%%%%%%%%%%%%%%%%%%%%%%%%%%%%%%%%%%%%%%%%%%%%

Note that double occupancy of the dot (even in a virtual state) is
forbidden since we have assumed $U=\infty$. All processes are
elastic in the sense that the energy of the QD has not changed
between its initial and final state. In particular, process (iii)
is elastic in the sense that the energy of the QD has not changed.
It is incoherent, though, since the spin in the QD has flipped (
and it is therefore possible to determine  that the electron under
study went through the interferometer's arm with the QD and not
through the free reference arm). Such a process will then
contribute to the total current but not to the flux-sensitive
component thereof, independent of the specific details of the AB
interferometer. The observation that energy exchange is not
necessary for dephasing
 \cite{Altshuler82} and that the latter can take place through,
e.g., a spin flip of an external degree of freedom, has been made
early on \cite{Stern90}.  In our case the electrons in the QD
itself (and their spin) serve as the ``dephasing bath''
\cite{Mello}.

To evaluate the transmission probability, hence the conductance, at
finite temperatures, one needs to sum over contributions from different
energies. This results in
\begin{eqnarray}
\label{conductance_finite_kT}
  {\cal{G}}_{\rm linear-response} \equiv
  {\partial I\over \partial V}\bigg|_{V=0} =
    - {e^2\over h}\sum_\sigma \int
    d \omega \,  T_{tr}^{\sigma}(\omega) f'(\omega) \, .
\end{eqnarray}
Here  $T_{tr}^{\sigma}(\omega)$ denotes the transmission
probability for an incoming electron of spin $\sigma$ and
frequency   $\omega$ while $f$ is the Fermi-Dirac distribution
function.

The transmission through a single-level QD  appearing in
Eq.~(\ref{conductance_finite_kT}) can be obtained
\cite{Meir92,Koenig96,Koenig99} from
\begin{eqnarray}
   T_{tr (\rm dot)}^{\sigma}(\omega) =
-{2\Gamma_L \Gamma_R \over \Gamma} \,{\rm Im}\,
    G^{\rm ret}_\sigma(\omega) \, .
\end{eqnarray}
Here the dot's  Green function, given by
\begin{eqnarray}
\label{QD_GF}
G^{\rm ret}_\sigma (\omega)=1/(\omega-\epsilon + i\Gamma/2)\,\,,
\end{eqnarray}
is the  Fourier transform of $ -i\Theta (t) \langle \left\{
c_\sigma (t),c^\dagger_\sigma (0) \right\} \rangle$. For
cotunneling, the transmission probabilities of electrons with
energy $\omega$ near the Fermi level of the leads can also be
obtained by calculating the transition rate in second-order
perturbation theory and multiplying it with the probabilities
$P_\chi$ to find the system in the corresponding initial state
$\chi$. For an incoming electron with spin up the transmission
probabilities are $P_\chi \Gamma_L \Gamma_R {\rm
Re}[1/(\omega-\epsilon+i0^+)^2]$.  Here  $P_\chi$ are the
probabilities to find the system in the corresponding initial
states $\chi=0,\uparrow,\downarrow$ for case (i), (ii), and (iii),
respectively.   Since $P_0 +P_\uparrow +P_\downarrow =1$ and
$P_0+P_\sigma =1/[1+f(\epsilon)]$ in equilibrium, we find
\begin{eqnarray}
\label{total_transmission_QD}
T_{tr (\rm dot)}^{\sigma}(\omega) = T_{tr (\rm dot, coh)}^{\sigma}(\omega) +
T_{tr (\rm dot, incoh)}^{\sigma}(\omega)
\end{eqnarray}
with \cite{comment_reg}

\begin{eqnarray}&&
\label{cotunneling_total,cotunneling_coherent}
%\label{cotunnelingtotal}
T_{tr (\rm dot)}^{\sigma}(\omega)={\rm
Re} \, {\Gamma_L \Gamma_R \over (\omega - \epsilon + i0^+)^2}
\nonumber \, ,
\\&& T_{tr {\rm dot, con}}^{\sigma}(\omega)= {T_{tr({\rm
dot})}^{\sigma}(\omega) \over {1+f(\epsilon)}} \, .
% \label{cotunneling_coherent}
\end{eqnarray}

The (evidently coherent) transmission amplitude through the QD can be defined
in the following way
\begin{eqnarray}
\label{GF_transmission_amplitude}
   t_{(\rm dot)}^{\sigma}(\omega) = i \sqrt{\Gamma_L \Gamma_R} \,
    G^{\rm ret}_\sigma (\omega) \, .
\end{eqnarray}
We now show that Eq.~(\ref{GF_transmission_amplitude}) is {\it not} a good
building block for calculating the transmission {\it probability} for
interacting  QDs.
Employing Eq.~(\ref{QD_GF}) we obtain
\begin{eqnarray}
\label{QD_transmission_amplitude} t_\sigma^{\rm dot}(\omega)= {i
(P_0 +P_\uparrow) \sqrt{\Gamma_L \Gamma_R}\over (\omega-\epsilon +
i0^+)}\, \, ,
\end{eqnarray}
from which it follows that
\begin{eqnarray}
\label{QD_transmission_amplitude_square}
|t_\sigma^{\rm dot}(\omega)|^2 =
 {T_{tr (\rm dot)}^{\sigma}(\omega)  \over [1+f(\epsilon)]^2}\,\,.
\end{eqnarray}
The latter equation does not yield the total transmission through
the dot, nor does it represent the  contribution to the
transmission probability arising from the coherent part of the
transmission.

We have thus demonstrated that the the presence of electron-electron interactions
gives rise to spin-flip processes, i.e. to incoherent transmission channels,
hence to the breakdown of  Eq.~(\ref{Landauer});
 there is no direct
physical meaning of the expression $|t_\sigma^{\rm dot}(\omega)|^2|$.

\section{How to break  phase locking}
\label{Phase_locking}
 The phenomenon of phase locking has been
found \cite{GIA1,GIA2} and recognized \cite{BIA} in the early days
of Mesoscopics. Experiments on small normal systems (simply and
multiply connected)  revealed  magnetoconductance which was
asymmetric under reversal of the direction of the magnetic field
 \cite{asymmetric_MR}. It has been then realized theoretically that the
magnetoresistance can be asymmetric  in the presence of an AB
flux. Following  some controversy  B\"uttiker has proposed a
multi-terminal effective circuit framework which captures the
essential symmetries of the problem. This approach is, in
principle, generalizable to a many-body interacting system coupled
to external terminals of independent electron gas.

Below we provide a brief review of B\"uttiker's approach
\cite{Buettiker86} (to be contrasted with the approach outlined
e.g. in Section \ref{Asymmetry}). We then make a few  comments
concerning the breakdown of the two-terminal phase locking
mentioned earlier.  Consider the four-terminal system depicted in
Fig.~\ref{fig3}.
%
%\begin{figure}
%\centerline{\includegraphics[width=8cm]{fig3.eps}}
%\caption{Four-terminal mesoscopic conductor
%         threaded by an AB flux.}
%\label{4_terminal}
%\end{figure}
%
The four terminals are connected to voltage sources (``electron
reservoirs'') of chemical potentials $\mu_i, i=1,...,4$. The
reservoirs serve both as a source and a sink of carriers and
energy. They possess the following properties: At zero temperature
they feed the leads with carriers up to the chemical potential
$\mu_i$. At finite temperatures they feed the leads at all
energies, weighted by the Fermi-Dirac function of the
corresponding temperature and chemical potential. Each carrier
coming from the lead and reaching the reservoir is absorbed by the
reservoir irrespectively of its phase and energy. We first assume
that the terminals connecting the respective reservoirs to the
system are strictly one-dimensional. This means that at each
terminal there are two running states at the Fermi energy, one
with  positive  velocity (away from the reservoir) and the other
with negative velocity. At this point we ignore interactions (e.g.
electron-electron) or any inelastic processes within the system
(inelastic relaxation and dephasing processes take place in the
reservoirs). The electrons are scattered elastically in the
system. We next assign scattering {\it probabilities}
$\{T_{ij}(\varphi)\}$ for carriers outgoing from terminal $j$ to
be transmitted into terminal $i$; we also use the notation
$\{R_{ii}(\varphi)\}$ to denote reflection probabilities from $i$
to $i$. It is clear that
\begin{eqnarray}
\label{symmetry_of_R_T}
&&T_{ij}(\varphi) = T_{ji}(-\varphi)\,\, ,
\nonumber \\&& R_{ii}(\varphi) = R_{ii}(\varphi) \, .
\end{eqnarray}
These relations can be easily verified employing equations akin to
Eqs.~(\ref{partial_amplitude}) and (\ref{total_reflection}).  As
we are interested in linear response, the differences among the
various $\{\mu_i\}$ are small, rendering the transmission and
reflection probabilities $\{T_{ij}(\varphi)\}$ and
$\{R_{ii}(\varphi)\}$ energy independent.

Straightforward algebra leads to the following expression for
the current in the $i${\it -th} lead
\begin{eqnarray}
\label{current_in_lead} I_i =  {e \over h} \left[(1-R_{ii})
\mu_i-\sum_{j \neq i} T_{ij} \mu_j \right]  \, .
\end{eqnarray}

Particle conservation, $R_{ii} +  \sum_{j \neq i} T_{ij} = 1$,
implies that  Eq.~(\ref{current_in_lead}) is independent of the
choice of the reference potential ($\mu=0$).

Let us first consider an arrangement \cite{Buettiker86,Casimir}
where the currents satisfy $I_1=-I_3$ and $I_2=-I_4$. Inserting
into  Eq.~(\ref{current_in_lead}) we obtain
\begin{eqnarray}
\label{chis}
 I_1 = \chi_{11} (V_1-V_3) - \chi_{12} (V_2-V_4) \, ,\\
 I_2 = -\chi_{21} (V_1-V_3) + \chi_{22} (V_2-V_4) \, ,
\end{eqnarray}
where $V_i = \mu_i / e$. B\"uttiker then found the following
expressions for the generalized conductances of  Eq.~(\ref{chis}):
\begin{eqnarray}&&
\label{chis_Ts}
 \chi_{11} ={e^2 \over h}\left[(1 - R_{11})S
-(T_{14} + T_{12})(T_{14} + T_{21})\right]/S  \, ,\\&&
\chi_{12} =
{e^2 \over h}(T_{12} T_{34} - T_{14} T_{32})/S  \, ,
\\&&
\chi_{21} = {e^2 \over h}(T_{21} T_{43} - T_{23} T_{41})/S  \, ,
\\&&
\chi_{22} = {e^2 \over h}\left[(1 - R_{22})S -(T_{21} +
T_{23})(T_{32} + T_{12})\right]/S \, ,
\end{eqnarray}
where
\begin{eqnarray}
\label{S}
 S = T_{12} + T_{14} + T_{32} + T_{34} = T_{21} + T_{41} + T_{23} + T_{43} \, .
\end{eqnarray}
The last equality  is obtained by connecting terminals $1$ and $3$ together, and
similarly terminals $2$ and $4$. One then obtains  a two-terminal geometry,
for which it is clear that the transmission $(1,3) \rightarrow (2,4)$ is equal
to the transmission in the reverse direction, $(2,4) \rightarrow (1,3)$.
Employing  Eq.~(\ref{symmetry_of_R_T}) one obtains
\begin{eqnarray}&&
\label{symmetry_of_chis}
 \chi_{11}(\varphi) = \chi_{11}(-\varphi) \, ,\\&&
 \chi_{22}(\varphi) = \chi_{22}(-\varphi) \, ,\\&&
 \chi_{12}(\varphi) = \chi_{21}(-\varphi) \, .
\end{eqnarray}
Eq.~(\ref{symmetry_of_chis})   establishes the Onsager relation
for this circuit.

We next  select the current source and
drain to  be terminals $1$ and $3$ respectively, and assume that
$2$ and $4$ are potentiometer terminals, implying that the
voltages $V_2, V_4$ are measures under the condition $I_2=I_4=0$. One
can now define {\it a} four-terminal conductance
\begin{eqnarray}
\label{G_{1324}} {\cal G}_{13,24} = {1 \over 2}{{I_1-I_3} \over
{V_2-V_4}} \, .
\end{eqnarray}
(We have used $I_1={1 \over 2}(I_1-I_3)$ to cast ${\cal
G}_{13,24}$ in a symmetric form). From Eqs.~(\ref{chis}),
(\ref{chis_Ts}) and (\ref{symmetry_of_chis}) one obtains
\begin{eqnarray}
\label{G_1324_chis} {\cal G}_{13,24}={{\chi_{11} \chi_{22} -
\chi_{12} \chi_{21}} \over {\chi_{21}}}\, .
\end{eqnarray}
Since $\chi_{21}$ is not symmetric in $\varphi$, ${\cal
G}_{13,24}$ turns out to be asymmetric as well (although the
Onsager relations are clearly satisfied \cite{Buettiker86}).

We now face the following paradox. Consider a two-terminal AB
interferometer. Let us now embed a ``conducting island'' in each
of the interferometer's arms.
%as depicted in Fig.~\ref{2_islands}.
Since this is a two-terminal setup, we expect phase-locking to
hold. On the other hand, once we make the embedded islands
sufficiently large, we may expect that eventually they would
represent electron reservoirs (as if we have added two extra
terminals to the circuit), leading perhaps to the breakdown of
phase locking? If this is indeed the case -- what is the
characteristic island size where this breakdown can be observed?
We first call attention to the fact that, as was discussed above,
the mere introduction of inelastic or phase breaking processes
cannot lead to the breakdown of phase locking.
%
%\begin{figure}
%\centerline{\includegraphics[width=8cm]{fig3.eps}}
%\caption{A two-terminal AB interferometer with two
%         conducting islands embedded in  each arms.}
%\label{2_islands}
%\end{figure}
%
To address these questions we note \cite{Feldman_Gefen} that in
the context of four-  ( or multi-) terminal geometries it is
possible to define other quantities whose dimension is
conductance. One quantity of interest is
\begin{eqnarray}
\label{G_1313} {\cal G}_{13,13}={1 \over
2}{{I_1-I_3}\over{V_1-V_3}} \, .
\end{eqnarray}
Similarly we define
\begin{eqnarray}
\label{G_1313'} {\cal G}_{13,13}'={{I_1}\over{V_1-V_3}} \, ,
\end{eqnarray}
and
\begin{eqnarray}
\label{G_1313''} {\cal G}_{13,13}''=-{{I_3} \over {V_1-V_3}} \, .
\end{eqnarray}
It is clear that as long as $I_2=I_4=0$,  ${\cal G}_{13,13}={\cal
G}_{13,13}'={\cal G}_{13,13}''$, and phase locking follows
immediately (we note that in this case the only efffect of the
extra terminals is to modify the effective elastic and inelastic
scattering rate in the two-terminal interferometer).

We now modify $V_i$ slightly into $V'_i =V_i+ \delta V_i,i=2,4$,
such that  there is  small current flowing out of terminal $2$
while the current through terminal $4$ is still zero. Then
$I_3=-(I_1+I_2)\ne -I_1$. and the definition of conductance
$G_{13,13}$ becomes ambiguous: $G_{13,13}=1/2(I_1-I_3)
/(V_1-V_3)\ne G_{13,13}'=I_1/(V_1-V_3)\ne
G_{13,13}''=-I_3/(V_1-V_3)$.   A simple calculation yields
\begin{eqnarray}&&
\label{1} G_{13,13}=
(\chi_{11}\chi_{22}-\chi_{12}\chi_{21})/\chi_{22}-\chi_{12}
I_2/[\chi_{22}(V_1-V_3)]=\nonumber \\&&
A(\Phi)+B(\Phi)I_2/(V_1-V_3) \,\,.
\end{eqnarray}
By Eq.~(\ref{symmetry_of_chis}) the first term in Eq. (\ref{1}) is
invariant under the transformation $\Phi\rightarrow -\Phi$.
However, the second term which is nonzero in the presence of $I_2$
is not invariant under this transformation. The condition for the
extremum of the conductance ( as function of the flux) is
$A'(\Phi)+I_2 B'(\Phi)/(V_1-V_3)=0$. Since $A'(0)=0$, we estimate
the derivative $A'(\Phi)$ as $A''(0)\Phi$ at small $\Phi$. For
small $I_2$ this gives for the extremum $\Phi_{\rm extr}=-I_2
B'(0)/[A''(0)(V_1-V_3)]$. The flux $\Phi_{\rm extr}$ is nothing
else but $-\alpha \Phi_0 / 2 \pi$, where $\alpha$ is the orbital
phase introduced above.

The above observations can be used for a systematic study of a
"gradual breaking" of phase locking, characteristic of
two-terminal geometries. A study of the breakdown of phase locking
in connection  the loss of unitarity may also be found in
Ref.~\cite{Aharony}.

We finally note that, in a sense,  phase-locking may be broken
even in a two-terminal geometry  if the applied magnetic flux is
not purely of a Aharonov-Bohm type. Once the magnetic flux
penetrates into the arms, different semi-classical trajectories of
the electron's path will enclose different amount of flux, and the
strict periodicity in $\Phi$ is broken. It follows that while
${\cal G}(\varphi)={\cal G}(-\varphi)$ still holds, the relation
${\cal G}(\varphi=n+\delta) ={\cal G}(\varphi=n-\delta)$ is
broken.

\section{The Dilemma of the transmission phase}
\label{Transmission_phase}

The discussion in Section (\ref{AB_general}) shows that in general
there is a coherent component of the electron transmitted through
a QD. It is therefore legitimate to ask what the {\it phase}
associated with the transmitted amplitude is. This is a
particularly interesting issue since the role of e-e interactions
in a low (zero) dimensional system, i.e. a QD, is expected to be
enhanced. This question has indeed been taken up by
experimentalists. It is clear that to obtain information about
quantum phase one needs to resort to interferometry experiments.
Typically the setup of such experiments consists of a 2DEG AB
interferometer. The latter includes a $G_aA_s/AlG_aA_s$
heterostructure QD embedded in one of its arms, and a free arm
which serves as reference. The QD is manipulated by varying its
gate voltage. In addition one controls the AB flux, the strength
of the dot-lead coupling and the temperature. The first
measurement (Ref.{\cite{Yacoby}), employing a two-terminal setup,
produced only limited information, having to do with the
phenomenon of {\it phase locking}. Later open-geometry experiments
(Ref.{\cite{Schuster}) yielded the flux dependent component of the
conductance of an AB interferometer ${\cal G}^{AB}(\varphi)$, as
function of the parameter $V_G$ (the gate voltage). This can be
written as
\begin{eqnarray}
\label{eq1} {\cal
G}^{AB}(\varphi,V_G)=A(V_G)g^{AB}(2\pi\varphi+\alpha(V_G)),
\end{eqnarray}
where $g^{AB}$ is a periodic function of $\varphi$. The prefactor
$A(V_G)$ is expected to be large (small) for values of $V_G$ that
correspond to Coulomb peaks (conductance valleys). Recalling
Eq.~(\ref{ltp}) for non-interacting electrons (in the double-slit
geometry alluded to above only the first harmonic in the AB flux
appears), it is tempting to draw an analogy between $\alpha(V_G)$
of Eq.~(\ref{Landauer})  and the orbital phase of
Eq.~(\ref{relative_AB_phase}). If the orbital part of the
transmission phase through the reference arm, $\alpha_1$, is
insensitive to the gate voltage, one may be motivated to refer to
$\alpha(V_G)$ (up to the constant $\alpha_1$), as the transmission
phase through the QD. In doing so we stress again that for
interacting electrons the transmission probability is {\it not}
given by the square of the transmission amplitude,
Eq.~(\ref{Landauer}).

Thus, in a strict sense, the naive interpretation of $\alpha(V_G)$
outlined above is wrong. One, however, notes that the flux
dependent part of the transmission {\it probability} is given by
\cite{Koenig01}
\begin{eqnarray}
\label{eq2}
T_2^{AB}(\omega)=2\sqrt{\Gamma_L\Gamma_R}|t^{\text{ref}}|\cos(2\pi\varphi)\text{Re}
G^{\text{ret}}(\omega)
\end{eqnarray}
for a two-terminal geometry and
\begin{eqnarray}
\label{eq3}
T^{AB}_{\text{open}}(\omega)=2\sqrt{\Gamma_L\Gamma_R}|t^{\text{ref}}|\text{Re}
|e^{-i(2\pi\varphi + \alpha)}G^{\text{ret}}(\omega)|
\end{eqnarray}
for an open geometry. Here $\Gamma_L,\Gamma_R$ are the dot-lead
couplings (on the left and right side respectively);
$t^{\text{ref}}$ is the transmission amplitude through the
reference arm alone); $G^{\text{ref}}$ is the retarded Green
function of the coupled dot and $\omega$ is the incident's
electron energy measured from the Fermi energy. The expression for
$T_{\text{open}}^{AB}(\omega)$ is analogous to the interference
term in Eq.~(\ref{ltp}) (roughly speaking, the left-right Green
function through the dot replaces the transmission amplitude
 through the dot, $t_2$). It is this fact that justifies referring
to $\alpha(V_G)$ as the transmission phase.

Typical parameters for the interferometry circuit are $U\approx
500\mu EV,\ k_BT\approx 10\mu eV$ and $\varepsilon_F\approx
10meV$. The number of electrons in the dot $N_{el}\approx
200-500$. The mean level spacing $\Delta\approx
\varepsilon_F/N_{el}\approx 50-150\mu eV$, hence $T\le\Delta$. The
dot-lead coupling $\Gamma\approx 1\mu eV<T$. This set of
parameters refers to the large QD experiments of
Refs.\cite{Yacoby} and \cite{Schuster}. In recent experiments
\cite{Ji1}, \cite{Ji2}  smaller QDs were used in order to
facilitate probing of Kondo physics. Our present review does not
include discussion of this limit.

%%%%%%%%%%%%%%%%%%%%%%%%%%%%%%%%%%%%%%%%%%%%%%%%%%%%%%
%%%%%%%%%%%%%%%%%%%%%%%%%%%%%%%%%%%%%%%%%%%%%%%%%%%%%%
%%%%%%%%%%% figphase             %%%%%%%%%%%%%%%%%%%%%
\begin{figure}
\epsfxsize = 0.8 \textwidth\centerline{\epsfbox{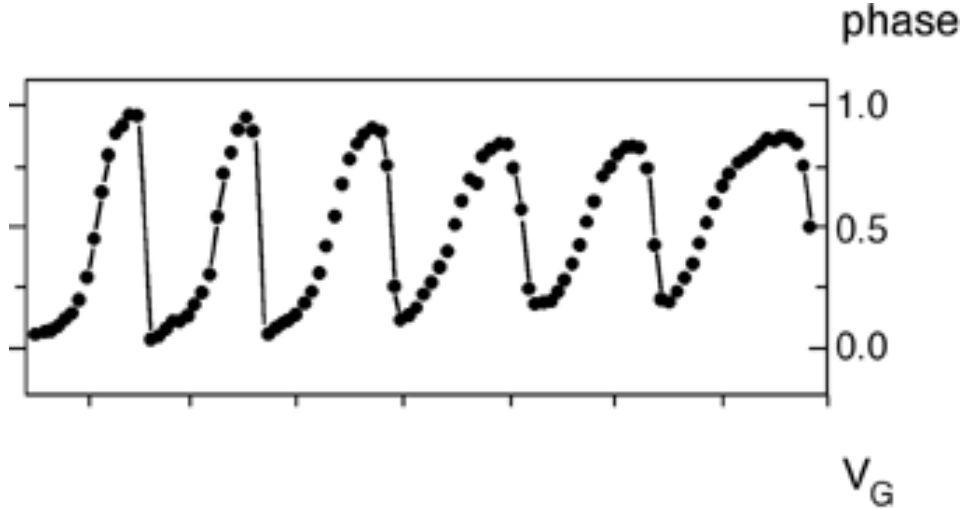}}
\caption{The evolution of the transmission phase $\alpha(V_G)$
from Ref. \cite{Schuster}} \label{figphase}
\end{figure}
%%%%%%%%%%%%%%%%%%%%%%%%%%%%%%%%%%%%%%%%%%%%%%%%%%%%%%
%%%%%%%%%%%%%%%%%%%%%%%%%%%%%%%%%%%%%%%%%%%%%%%%%%%%%%

The evolution of $\alpha(V_G)$ is shown in Fig.~\ref{figphase} as
function of the gate voltage. The latter is swept across
consecutive Coulomb resonances. The phase $\alpha$ increases by
$\pi$ as $V_G$ is swept across a Coulomb peak [deviations from
$\pi$ are, presumably, due to the fact that the induced resonances
are not entirely independent: the ratio $\Delta/\Gamma$ is not
sufficiently large]. This comes as no surprise and can easily be
accounted  for if one represents each individual resonance by a
Breit-Wigner Lorentzian. The width of the $\pi$-step in $\alpha$
at resonance  is of the order of the resonance width. At low
temperatures the latter is denoted by $\tilde\Gamma$ (in Section
\ref{Width}  we comment on the differences between the bare
(golden rule) level width $\Gamma$ and $\tilde \Gamma$), and
remains so as long as $k_3T<<\tilde\Gamma$. In the conductance
valley (i.e., between consecutive Coulomb peaks), the phase
appears to drop rather sharply by $\pi$, rendering the phase
evolution over a single period (in $V_G$) of Coulomb oscillations
0. These valley-to-valley correlations in the transmission phase
have been observed repeatedly in a number of measurements,
spanning up to 12 consecutive peaks in a single measurements.

This remarkable result soon attracted the attention of theorists,
stimulating a large number of papers attempting to explain this
phenomenon. I shall not try to present a comprehensive overview of
these works. Instead I will mention the main approaches and state
to what extent the problem still remains open.

The first point to note in this context is that these transmission
phase correlations cannot be accounted for by independent electron
theories \cite{Berkovits98}. Indeed, for a non-interacting
``one-dimensional QD'' (a segment of a one-dimensional wire of
length $L$ bounded by two potential barriers, whose respective
transmission and reflection amplitudes from left (right) are
$t_1,r_1,t_2,r_2$ $(t'_1,r'_1,t'_2,2'_2)$) the transmission
amplitude for an incoming electron of wave-number $k$ is
\begin{eqnarray}
\label{eq4} t_{1dQD}=\frac{t_1t_2e^{ikL}}{1-r'_1r_2e^{i2kL}}.
\end{eqnarray}
It is easy to determine that as one varies the incoming electron's
energy (or, alternatively, the base potential of the
one-dimensional QD) the phase of $t$ varies by $\pi$ across a
transmission resonance (i.e., across the energy of a quasi-bound
state). This, however, is not accompanied by any phase lapse
between resonances. The overall change of the transmission phase
over a single period (as we increase the energy of the incoming
electrons from, say, just below the $n$-{\it th} resonance to just
below the $(n+1)$-{\it st} resonance) is therefore $\pi$, in stark
contradistinction with the experimental data. This
change-by-$\pi$-per-period is intimately connected with the fact
that the sign $(p(n))$ of the product of the coupling matrix
elements of the $n$-th wave function of the dot to the left and to
the right leads alternates with $n$ \cite{Berkovits98}  (We
consider here a system which possesses time reversal symmetry
(e.g., no magnetic field)). The single particle wave functions of
the uncoupled dot can thus be chosen to be real; by an appropriate
gauge the dot-lead matrix elements can also be made real. The sign
of the latter \cite{Berkovits98}, \cite{Duke69} is that of the
derivative of the component of the wave function normal to the
dot-lead interface, cf. Ref \cite{Bardeen61}. For a two- or
three-dimensional non-interacting dots $p(n)$ is geometry
dependent and, in general, does not show any robust
$n$-independence to account for the experimental data.
Furthermore, for chaotically shaped or diffusively disordered QDs,
the signs of the derivative of consecutive single-particle wave
functions (near the lead), hence $p(n)$, are (to leading order)
uncorrelated (unlike spectral properties). It is thus evident that
we need to go beyond the independent particle framework.

Some of the early attempts to resolve the transmission-phase
correlation effect \cite{Yeyati95} (see also \cite{Hofstetter})
relied on the Friedel sum rule \cite{Friedel52,Langer61} which
provides for a relation between phase and charge. One needs,
though, to call attention to the fact that the Friedel sum rule
deals with the {\it scattering matrix} (rather than the
transmission matrix). It relates the total charge displaced in the
field of a fixed impurity (e.g., the charge added to a quantum
dot) to the scattering by that impurity of a free electron at the
Fermi momentum $k_F$. The number of displaced electrons, $N_D$, is
given by
\begin{eqnarray}
\label{eq5} N_D = \frac{1}{\pi}\sum_{l,m_l,m_s}\delta_{l,m_l,m_s},
\end{eqnarray}
where the sum runs over the scattering phases $\delta$ with
$(l,m_l),m_s$ being the angular momentum and the spin quantum
numbers. More generally one can write
\begin{eqnarray}
\label{eq6} N_D = \frac{1}{2\pi i}T_r \ln S(\mu)
\end{eqnarray}
where $S(\mu)$ is the scattering matrix for single-particle-like
excitations at the chemical potential $\mu$. Consider for a moment
the scattering of non-interacting electrons in one-dimension
\cite{Anderson80}. (We assume that the scattering is spin
independent, hence suppress the spin index). This two-channel
problem is described in terms of a $2\times 2$ matrix

\begin{eqnarray}
\label{eq7} S = \left(\matrix{ r & t' \cr t & r' \cr}\right)
\end{eqnarray}

whose eigenvalues are $e^{i\theta_l}, l=1,2$ (primed quantities
refer to reflection and transmission coefficients for electrons
impinging on the scatterer from the right). The one-dimensional
version of the Friedel sum rule asserts that \cite{Anderson80}
\begin{eqnarray}
\label{eq8}
n(\varepsilon)=\frac{1}{2\pi}\frac{\delta(\theta_1+\theta_2)}{\delta
\varepsilon}\, \, ,
\end{eqnarray}
where $n(\varepsilon)$ is the density of states contained in the
scatterer. It can also be shown that for the 1d case the
transmission amplitude, parameterized as $t=|t|e^{i\alpha}$, leads
to the relation
\begin{eqnarray}
\label{eq9} \frac{\theta_1+\theta_2}{2}=\alpha +
\frac{\pi}{2}\qquad (1d) \, \,.
\end{eqnarray}

Therefore, specifically for the 1d case, the Friedel sum rule can
be expressed, e.g., through Eq.~(\ref{eq9}), with the transmission
phase $\alpha$ replacing the scattering phases. This, however, is
not a general theorem concerning the transmission phase.
Furthermore, there is no reason why the total charge accumulated
at the scatterer (the QD and the leads near it) over a Coulomb
period should be an integer (let alone 0, as is required if the
total charge of the scattering phase were to be 0).

The theoretical effort addressing the correlations in the
transmission phase could be divided, in large part, into two
approaches.

The first school of thought maintains that there are one or few
dot levels which are particularly strongly coupled to the leads.
Such a strongly coupled level will dominate a number of
consecutive transmission peaks. This would imply that successive
resonances are dominated by the same tunneling matrix elements. In
other words, it is practically the same level which keeps
repeating at consecutive resonances, leading to transmission phase
correlations over successive Coulomb periods.

As a specific example one may invoke a model QD whose Hamiltonian
is made primarily of an integrable part (its eigenstates are
products of longitudinal and transverse modes). The subset of
states possessing high longitudinal quantum numbers defines the
strongly coupled levels. The location of the gates is chosen in
such a way that the energies $\{ E^0_\alpha \}$ of the strongly
coupled states, $\{ F_\alpha \}$, are weakly dependent on the gate
voltage. We now switch on a small non-integrable term of the
potential. This leads to avoided level crossing (of the original,
"bare" levels), as function of $V_G$. As is demonstrated in
Fig.(\ref{figfloating})  an actual single-particle level,
$\psi_n$, (plotted as function of $V_G$), is now made piecewise of
strongly (flat) and weakly (steep) coupled bare states (F's and
S's). In particular, a given $F_\alpha$ will be equal to $\psi_n$
for a certain window of $V_G$, to $\psi_{n+}$ for the next
interval of $V_G$ etc. It so happens that as the levels
$n,n+1,n+2,\ldots$ cross successively into the Fermi sea, what
used to be the (bare) level $F_\alpha$ will keep ``floating'' over
the Fermi level, dominating successive Coulomb resonances.

 While this picture \cite{Hackenbroich97,Oreg97}
provides a correlation-generating mechanism, it has a couple of
nagging weaknesses. Firstly, for the level $F_\alpha$ to keep
"hovering" just above the Fermi energy, an (approximate)
commensurability condition is required between intervals (in
$V_G$) of consecutive avoided crossings and intervals (in $V_G$)
over which an additional electron is added to the QD (Coulomb
period). This imposes rather stringent constraints on the geometry
and the confining potential. Secondly, this picture assumes that
the dot's levels (varying as function of $V_G$) are made piecewise
of the original bare levels. The latter, other than at value of
$V_G$ that correspond to avoided crossing, do not mix, implying
that the effect of the non-integrable term in the Hamiltonian is
weak. Some of the QDs studied in the experiments of the Weizmann
group might be indeed almost integrable, and the above approach
may be suited to describe the pertinent physics. On the whole,
though, this approach lacks the flavor of being generic, i.e.,
pertaining to chaotic or diffusive QDs where level mixing is
strong. As for the first reservation mentioned above the good news
is that both finite temperature \cite{Baltin99} or quantum
\cite{Silvestrov00} fluctuations render the ``hovering effect''
(i.e., the dominance of a single level $S_\alpha$ over a number of
consecutive Coulomb peaks) more robust.

\begin{figure}[t]
%\rule{5cm}{0.2mm}\hfill\rule{5cm}{0.2mm}
\vspace{8.5cm}
%\centerline{\psfig{figure=moriond_plot.ps,height=0.3in}}
%\rotatebox{-90}{\psfig{figure=moriond_plot.ps,height=0.3in}}
\includegraphics{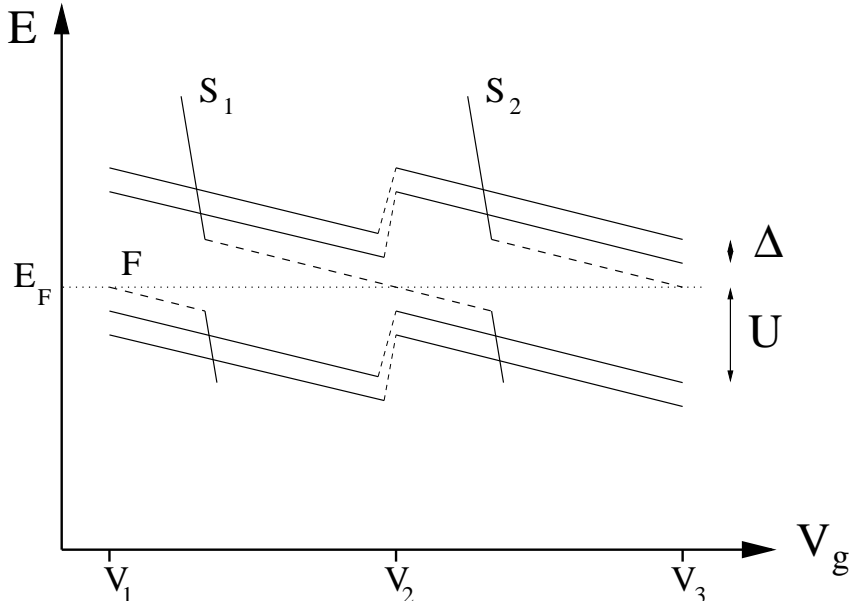} \caption{Avoided crossings with the steep
bare levels $S_1$ and $S_2$ cause the flat- and strongly coupled-
bare level $F$ (dashed) to stay close to the Fermi energy
$\varepsilon_F$. Resonances dominated by this level occur at gate
voltages $V_1,V_2,V_3$. There is a gap of magnitude $U$ between
the last occupied and the first empty level. From Ref.
\cite{Baltin99}} \label{figfloating}
%\rule{5cm}{0.2mm}\hfill\rule{5cm}{0.2mm}
%\vskip 9cm
%\caption{Interaction correction to the noise (solid line) and the
%"naive  expectation" (dashed line), see text for definitions.}
%\label{fig:radish}}
\end{figure}
%%%%%%%%%%%%%%%%%%%%%%%%%%%%%%%%%%%%%%%%%%%%%%%%
%%%%%%%%%%%%%%%%%%%%%%%%%%%%%%%%%%%%%%%%%%%%%%%%
%%%%%%%%%%% figfloating    %%%%%%%%%%%%%%%%%%%%%
%\begin{figure}
%\epsfxsize = 0.5 \textwidth \epsfbox{tmuna2.eps}\caption{Avoided
%crossings with the steep bare levels $S_1$ and
% $S_2$ cause the flat- and strongly coupled- bare level $F$
% (dashed) to stay close to the Fermi energy $\varepsilon_F$.
% Resonances dominated by this level occur at gate voltages
% $V_1,V_2,V_3$. There is a gap of magnitude $U$ between the last
% occupied and the first empty level. From Ref. \cite{Baltin99}}
%\label{figfloating}
%\end{figure}
%%%%%%%%%%%%%%%%%%%%%%%%%%%%%%%%%%%%%%%%%%%%%%%%%%%%%%
%%%%%%%%%%%%%%%%%%%%%%%%%%%%%%%%%%%%%%%%%%%%%%%%%%%%%%

The second school of thought addressing the phase correlation
effect takes the opposite point of view. Rather than a single,
particularly strongly coupled level, dominating the transmission
phase over a wide interval of $V_G$, here we rely on the fact that
the coherent transmission in the ``conductance valley'' (i.e.,
between Coulomb peaks) is mostly due to the process of elastic
cotunneling. A parametrically large number of levels (of order
$U/\Delta$) participate, each making a small (in magnitude) and
random (in magnitude and phase) contribution to the transmission
amplitude through the QD. Shifting the gate voltage to the next
valley, these are almost the very same dot's levels that
contribute, leading (with a high probability) to the same phase of
the transmission amplitude. This is the mechanism behind the
valley-to-valley correlations. Detailed analysis of the phase
evolution following this picture is presented in Ref.
\cite{Baltin99a}

\section{Asymmetry of the interference signal}
\label{Asymmetry} Let us consider transport through the QD  away
from resonance. At temperatures higher than the Kondo temperature
this   is dominated by cotunneling (second order in $\Gamma$).
From  the discussion of Section \ref{AB_general} \cite{Koenig01}
it turns out that  such cotunneling effects give rise to an
asymmetry of the AB amplitude measured on either side of a Coulomb
conductance peak. To see this consider, for example, the two level
QD modelled by the Hamiltonian of Eq.~\ref{QD_Hamiltonian}. We
tune the gate voltage such that the Fermi energy is a distance
$\epsilon (\ll U)$ above (or below) the  Coulomb resonance
separating the $N_{el}=0$ from the $N_{el}=1$ valley ($N_{el}$ is
the mean number of electrons on the dot). For $\epsilon \gg
\Gamma, k_B T$ the dominating  cotunneling process on the
$N_{el}=0$ side of the conductance peak is the first (coherent)
process depicted in Fig.~\ref{figcotunneling}, while on the
$N_{el}=1$ side cotunneling is dominated by the  other  two
processes depicted in Fig.~\ref{figcotunneling}. These two
processes are of equal probability $\sim \Gamma_L \Gamma_R {\rm
Re}[1/(\omega-\epsilon+i0^+)^2]$. Only  one of them is coherent
(the second in that figure) while the other contributes to the
current through the QD but not to the flux dependent conductance
${\cal G}^{AB}(\varphi)$.  If we compare two values of $V_G$ on
either side of the conductance peak for which the flux-averaged
conductance is the same, we expect greater visibility on the
$N_{el}=0$ side, hence (cf. Eq.~\ref{visibility}) a larger AB
amplitude on that side. This would imply that the the AB amplitude
is asymmetric with respect to the total conductance (as a function
of $V_G$).

Our detailed analysis \cite{Koenig01}  reveals that this asymmetry
exists  both near  resonance (going to first order in $\Gamma$)
and in second order. Considering an AB interferometer with a QD in
one of the arms, and tuning  the transmission of the reference
arms to $|t_{\rm ref}| = \sqrt{\Gamma_L \Gamma_R} / |\epsilon|$
(to maximize the visibility) one finds for the total conductance
\begin{equation}
\label{cotunneling noninteracting}
   {\partial I^{\rm tot} \over \partial V}\bigg|_{V=0} = {4e^2\over h}
    {\Gamma_L \Gamma_R \over \epsilon^2} \left[
    1 - {\epsilon \over |\epsilon|} \cos \varphi \right]
\end{equation}
for the noninteracting case ($U=0$) and
\begin{equation}
\label{cotunneling interacting}
   {\partial I^{\rm tot} \over \partial V}\bigg|_{V=0} = 4 {e^2\over h}
    {\Gamma_L \Gamma_R \over \epsilon^2} \left[
    1 - {\epsilon \over |\epsilon|} {\cos \varphi \over 1+f(\epsilon)}
    \right]
\end{equation}
for $U=\infty$. This shows that cotunneling in the noninteracting
case is fully coherent (we can tune both $t_{\rm ref}$ and
$\varphi$ such that the total transmission probability -- hence
the conductance-- vanishes. In the interacting case spin-flip
processes are present which spoil coherence. This is described by
the asymmetry factor $1/[1+f(\epsilon)]$, in accordance with our
intuitive picture (in the above expressions the conductance peak
is at $\epsilon=0$.)

The above considerations can be generalized for a multilevel QD.
The asymmetry factor can be used to obtain information concerning
the total spin of the QD in a parameter regime away from the Kondo
limit.

In Fig.\ref{asymmetry_small_qd} and  Fig.\ref{asymmetry_large_qd}
we present unpublished data concerning the dependence on gate
voltage of both the  total conductance through the AB
interferometer and  and magnitude of the flux-modulated amplitude.
The curves presented in Fig.\ref{asymmetry_small_qd} agree
qualitatively with the above discussion: for the small QD data it
is possible to identify  and distinguish between Coulomb blockade
valleys (where the total number of electrons is presumably even)
and Kondo valleys ( odd number). The latter can be identified
through the very low temperature behavior of the conductance, not
shown here \cite{Ji1,Ji2}. The theoretical prediction is that the
peaks of the AB curve are shifted asymmetrically ( with respect to
the conductance peaks) away from the Kondo valleys, which is
indeed suggested by Fig.\ref{asymmetry_small_qd}. Surprisingly
enough this seems not to be the case for the large dot curve,
Fig.\ref{asymmetry_large_qd}, where the peaks of the AB amplitude
appear to be all shifted to the right of the conductance peaks.

%%%%%%%%%%%%%%%%%%%%%%%%%%%%%%%%%%%%%%%%%%%
%%%%%% figure(asymmetry_small_qd.eps) goes here %%%%%%
%%%%%%%%%%%%%%%%%%%%%%%%%%%%%%%%%%%%%%%%%%%%

\begin{figure}
\epsfxsize = 0.9 \textwidth
\centerline{\epsfbox{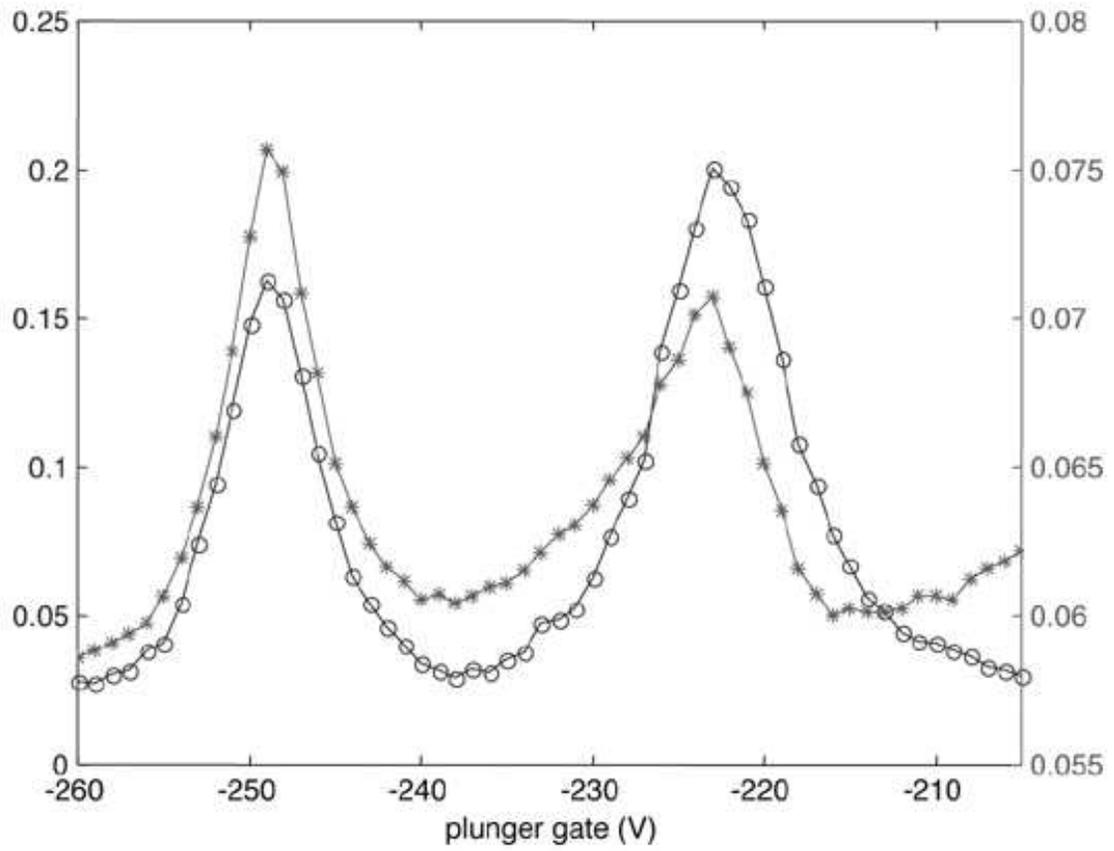}} \caption{Total
conductance(circles) and the AB amplitude (dots) as function of
$V_G$  for a single (small) QD interferometer. Data are courtesy
of Yang Ji and M. Heiblum} \label{asymmetry_small_qd}
\end{figure}
%%%%%%%%%%%%%%%%%%%%%%%%%%%%%%%%%%%%%%%%%%%%

%%%%%%%%%%%%%%%%%%%%%%%%%%%%%%%%%%%%%%%%%%%
%%%%%% figure(asymmetry_large_qd.eps) goes here %%%%%%
%%%%%%%%%%%%%%%%%%%%%%%%%%%%%%%%%%%%%%%%%%%%
%%%%%%%%%%%%%%%%%%%%%%%%%%%%%%%%%%%%%%%%%%%%
\begin{figure}
\epsfxsize = 0.9 \textwidth
\centerline{\epsfbox{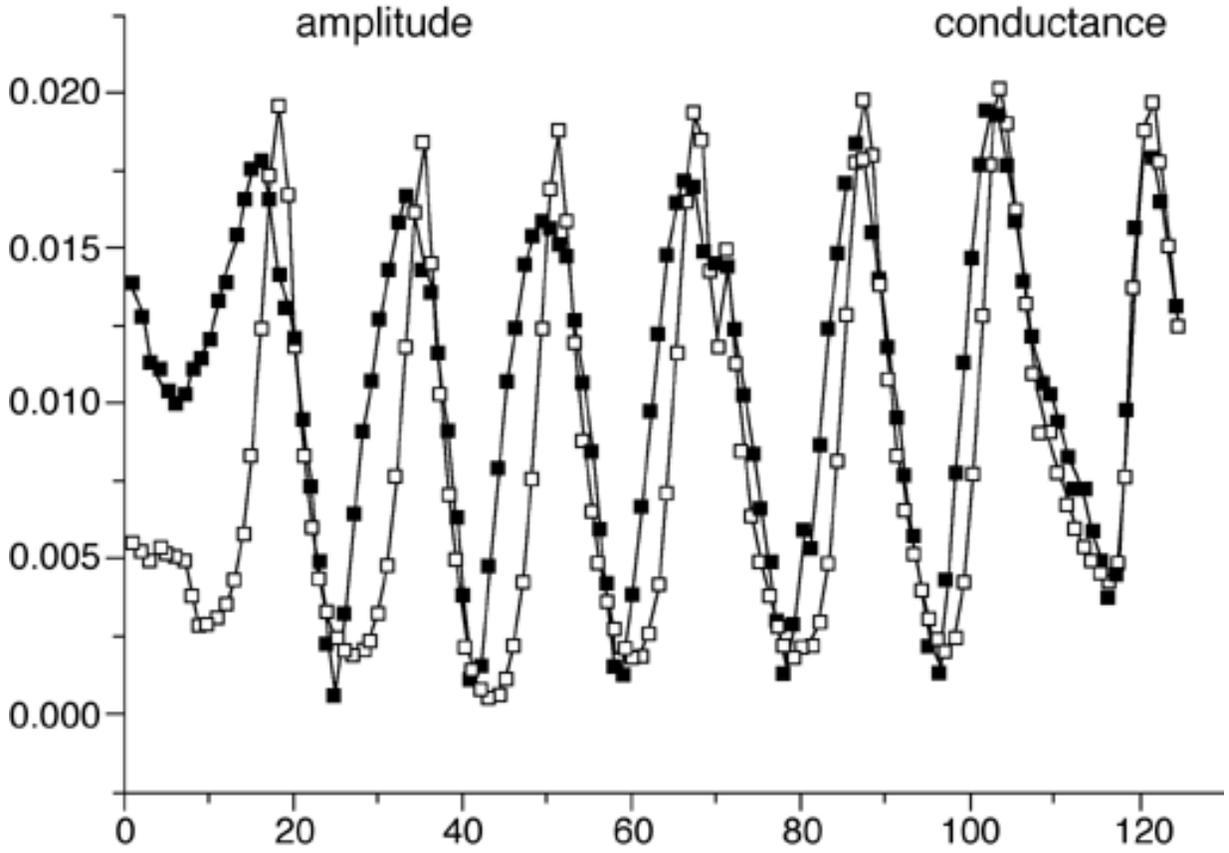}} \caption{Total
conductance (grey) and AB amplitude (black) as function of $V_G$
for a large QD interferometer. Data are courtesy of R. Schuster,
Yang Ji and M. Heiblum} \label{asymmetry_large_qd}
\end{figure}

One might note the shape of the conductance ( and the AB) peaks is
asymmetric as well. It was proposed that this is a manifestation
of the Fano resonance effect, cf. Refs.
\cite{Fano,Hist,Goeres,Bulka,Hofstetter,EAIL}.

\section{On the Width of the Resonance and the Phase Lapses}
\label{Width}

The evolution of the transmission phase discussed above presents
us with further dilemmas which have been pretty much ignored till
now. These concern with the widths of both the phase change by
$\pi$ at the Coulomb peaks and the phase change (again by $\pi$)
at the phase lapses. These widths are measured on the scale of
(the change of) the gate voltage $V_G$. The issue is yet far from
being resolved. Here we shall present the problems and add a few
comments.

Let us first consider the range of $V_G$ (near resonance) over
which the phase change by $\pi$ takes place. This is also the
width of the Coulomb peak. We first consider ``metallic dots'',
meaning that $k_BT>\Delta$. It is commonly accepted that (at least
for a multichannel dot-level coupling) the physics of the Coulomb
peaks is a function of the dimensionless dot-level conductance,
$g_{DL}$. We will assume that the couplings of the $QD$ to the
left and to the right leads are of comparable strengths. We also
note that $g_{DL}\sim\frac{\Gamma}{\Delta}$, where $\Gamma$ is the
(bare) golden-rule width of a dot's level. One should be careful
distinguishing $\Gamma$ from $\tilde\Gamma$, the latter being the
width of the Coulomb peak. In the weak coupling limit
\cite{Alhassid} $\tilde\Gamma\sim\max(\Gamma,k_BT)$ (for a
metallic dot $\tilde\Gamma$ is dominated by $k_BT$ in this
limit)\cite{Footnote3}.
 To get the
flavor of the dilemma involved, let us now replace the actual
Coulomb peaks by a periodic (in $V_G$) sequence of identical
Lorentzians
\begin{equation}
\label{eq10} \frac{d(\text{conductance})}{d(\text{energy})}=
\frac{e^2}{\pi h}\sum_m
\frac{\tilde\Gamma}{(V_G-mU)^2+\tilde\Gamma^2}
\end{equation}
The Fourier transform of the above expression (with respect to the
gate voltage) is
\begin{equation}
\label{weak} \frac{e^2}{2\pi h}\sum_n e^{-\frac{2\pi n\tilde
\Gamma}{U}} e^{i2\pi n V_G/U}.
\end{equation}
This behavior is characteristic of mesoscopic systems. For
observables which are periodic in some parameter, higher harmonics
are suppressed faster (as function of width, inelastic rate,
dephasing rate etc.)

Turning now our attention to the strong coupling limit one expects
that the periodic modulation (of the conductance, the derivative
of the particle number etc.) is all but suppressed. Only
exponentially small modulation survives (of order
$e^{-\frac{\pi^2}{8}g_{DL}}$) \cite{Nazarov,Kamenev}. In the
language of the above Fourier expansion this amounts to the first
harmonic being
\begin{equation}
\label{strong} e^{-\frac{\pi}{4}\frac{\Gamma}{\Delta}} e^{i2\pi n
V_G/U},
\end{equation}
(higher harmonics will be suppressed even further)
\cite{footnote_renormalization}.  The fact that the (exponentially
small) first harmonic dominates implies that the width of the
(exponentially small) ``Coulomb peak'' is $\sim U$. Comparing
Eqs.~(\ref{weak}) and (\ref{strong}) (at $\Gamma/\Delta \sim 1$)
reveals that in the vicinity of the weak-to-strong-coupling
crossover, $\tilde\Gamma$ changes dramatically  from
$\tilde\Gamma\sim k_BT$ (or $\Gamma$) to $\tilde\Gamma\sim U$ over
a rather small interval of $V_G$. We also note that
Eqs.~(\ref{weak}) and (\ref{strong}) cannot be reconciled within a
single parameter scaling theory \cite{single_parameter}. We stress
that at this point our considerations are rather qualitative. We
expect a similar fast crossover of $\tilde\Gamma$ (from the weak
to the strong coupling limit) with other quantities as well, e.g.
$d\langle N_{el}\rangle/d_{V_G}$, where $\langle N_{el}\rangle$ is
the expectation value of the number of electrons on the QD. We
also expect a fast crossover of $\tilde\Gamma$ in the discrete
level limit ($k_BT<\Delta$) as well.

Let us now turn our attention to the width of the phase lapses
that occur in the ``Coulomb valleys'', between Coulomb peaks. As
has been discussed earlier, a universally accepted theory for such
phase lapses is not yet available. Here we focus on another
interesting observation -- it appears that the experimentally
observed typical width of these phase lapses, $\Gamma_{PL}$, is
significantly smaller than that of the Coulomb peak,
$\tilde\Gamma$. Presently this is a qualitative observation, yet
to be backed up by a detailed study.

It is indeed a challenge to find a mechanism which provides for
$\Gamma_{PL}$ which is parametrically smaller that $\tilde\Gamma$.
In order to develop a feel as to what the difficulty is let us
consider a toy model which exhibits a phase lapse. This is a
spinless two-level QD coupled to two leads with the Hamiltonian
(cf. Eq.~\ref{QD_Hamiltonian})
\begin{equation}
\label{2level_QD} H=\sum \epsilon_{k,\alpha} \;
c^{\dagger}_{k,\alpha} c_{k,\alpha}+ \sum_j \epsilon_j
d^{\dagger}_jd_j+ \nonumber \\
+\sum_{k,\alpha,j}
\left[V_{\alpha,j}c^{\dagger}_{k,\alpha}d_{j}+h.c\right],
% \mathcal{H}=\sum \varepsilon_k c_{k,i}^+c_{k,i}
%+ \sum_{j=1,2}\tilde\varepsilon_jd_j^+ d_j + \sum_{k,i,j} V \left(
%c_{k,i}^+ d_j + h.c.\right).
%
\end{equation}
Here the operators $c_{k,\alpha}$ refer to the electronic states
in the leads ($i=L,R$) and the operators $d_1,d_2$ are associated
with the QD states. We note that each of the dot's level will
acquire a width $\Gamma$ due to its coupling to the leads (this
width may be further renormalized due to higher order dot-lead
tunnelling processes, c.f. Ref. \cite{KGS}). We have intentionally
chosen all dot-lead matrix elements to have the same sign. It can
be shown that this indeed leads to a phase lapse
\cite{Hackenbroich,OG}. Moreover, the original analysis, treating
the tunnelling into/from each dot level independently, results in
the phase lapse having a width of $\Gamma=2\pi V^2 \rho$ ($\rho$
is the density of status in the leads). However, our recent
analysis of Eq. \ref{2level_QD} \cite{silva} shows that the
tunnelling-induced coupling between the QD's levels must not be
neglected. Since the toy model at hand is of non-interacting
electrons, its analysis is straightforward. The single electron
Greens function is given by
\begin{equation}
\label{greens} G(w) = \left[w-\tilde{\mathcal{H}}\right]^{-1}
\end{equation}
where, in terms of the variables
$\tilde\varepsilon=-(\varepsilon_1+\varepsilon_2)/2$ and
$\delta_G=(\varepsilon_1-\varepsilon_2)/2$ the effective
Hamiltonian of the QD (in the Hilbert space of levels 1,2) is
%\begin{eqnarray}&&
%\label{eq11} \tilde{\mathcal{H}}= \left(\matrix{
%-\tilde\varepsilon + \delta\varepsilon-i\Gamma & -i\Gamma \cr
%-i\Gamma & -\tilde\varepsilon - \delta\varepsilon - i\Gamma
%\cr}\right).
%\end{eqnarray}

\begin{eqnarray}&&
\tilde{\mathcal{H}}=\left(\matrix{-\tilde\varepsilon +
\delta\varepsilon-i\Gamma & -i\Gamma \cr -i\Gamma
&-\tilde\varepsilon - \delta\varepsilon - i\Gamma \cr}\right).
\end{eqnarray}

One can readily find
\begin{eqnarray}&&
\label{eq12} G(w) = \frac{1}{D(w)}\left(\matrix{
w+\tilde\varepsilon + \delta\varepsilon + i\Gamma & -i\Gamma \cr
-i\Gamma & w + \tilde\varepsilon - \delta\varepsilon + i\Gamma
\cr} \right),
\end{eqnarray}
when the determinant
$D(w)=(w+\tilde\varepsilon)^2-(\delta\varepsilon)^2+2i\Gamma(w+\tilde\varepsilon)$.
The transmission amplitude from left to right can be written as
\begin{equation}
\label{t_2level} t(w) = \Gamma \sum_{i,j=1,2} G_{i,j}(w) =
\frac{2\Gamma}{D(w)} (w+\tilde\varepsilon).
\end{equation}
Varying the gate voltage amounts to varying $\tilde\varepsilon$
(leaving all other parameters unchanged). It is clear from Eq.
\ref{t_2level} that one can tune the gate voltage (hence
$\tilde\varepsilon$) to obtain an {\it exact} zero of the
transmission amplitude between two peaks. This occurs for
$w=-\tilde\varepsilon$. Sweeping $V_G$ around this point results
in a sign change of $t(w)$, hence a zero-width phase lapse
\cite{silva}. This phase lapse acquires a finite width at finite
temperatures, or when the hopping matrix elements assume
non-trivial relative phases. The above discussion (presented here
for a non-interacting QD) demonstrates that the physics
responsible for the width of phase lapses is quite different from
that applicable at resonances, and may indeed give rise to
parametrically narrow $\Gamma_{PL}$.

{\bf Acknowledgment}.\, I acknowledge useful discussions with  Y.
Imry, A. Kamenev and H.~A. Weidenm\"uller. This overview employs
results and relies on insights obtained in the course of my
present collaboration with D.~E. Feldman, J. K\"onig, Y. Oreg and
A. Silva. I am indebted to my colleagues Yang Ji, M. Heiblum and
R. Schuster for discussions concerning the experiments and for the
permission to present unpublished data. This work was supported by
the U.S.-Israel Binational Science Foundation, by the GIF, by the
Israel Science Foundation and by the Minerva Foundation.

%\end{article}


\begin{thebibliography}{99}
\bibitem{refGIA1}
Note that the phase locking symmetry is present in the results of ref. \cite{GIA1}.
\bibitem{Footnote2}
 In ref.\cite{Koenig01} it is shown why, in the
absence of electron-electron interaction in the dot, the amplitude
of spin-flip processes vanishes.
\bibitem{comment_reg}
The regularization $+i0^+$ is put here by hand, but can be derived
within a complete cotunneling theory \cite{Koenig_cotunneling}.
\bibitem{footnote_renormalization}
Note that while the height and the width of the Coulomb peak are
strongly renormalized in the strong coupling limit, this is not
the case with the peak-to-peak distance $U$. The latter might be
modified by trivial factors, involving the ratio of
self-capacitance/dot-gate capacitance.
\end{thebibliography}
\end{document}